\documentclass[12pt,showpacs,preprintnumbers,superscriptaddress,amsmath,amssymb,nofootinbib]{revtex4}

\usepackage{graphicx}
\usepackage{amsmath}
\usepackage{amssymb}
\usepackage{bm}
\usepackage{multirow}
\allowdisplaybreaks[1]

\newcommand{\nl}{\nonumber\\}
\renewcommand{\ol}{\overline}
\newcommand{\bea}{\begin{eqnarray}}
\newcommand{\eea}{\end{eqnarray}}

\newcommand\fverbdo{\egroup\medskip\noindent%
			\fbox{\unhbox\fverbbox}\ }
\newcommand\fverbit{\egroup\item[\fbox{\unhbox\fverbbox}]}
\newbox\fverbbox

\newcommand{\al}{\alpha}

\newcommand{\ga}{\gamma}
\newcommand{\la}{\lambda}
\newcommand{\si}{\sigma}

\newcommand{\De}{\Delta}
\newcommand{\Ga}{\Gamma}
\newcommand{\La}{\Lambda}
\newcommand{\Om}{\Omega}

\begin{document} 
\allowdisplaybreaks[2]

\title{Dark matter and muon $(g-2)$ in local $U(1)_{L_\mu-L_\tau}$-extended Ma Model}

\preprint{KIAS-P15054}

\author{Seungwon Baek}
\email{swbaek@kias.re.kr}
\affiliation{School of Physics, KIAS, 85 Hoegiro, Seoul 02455, Korea}



\begin{abstract}
We consider right-handed neutrino dark matter $N_1$ in local $U(1)_{L_\mu-L_\tau}$-extended Ma model.  With
the light $U(1)_{\mu-\tau}$ gauge boson ($m_{Z'} \sim {\cal O}(100)$ MeV) and small $U(1)_{\mu-\tau}$ gauge coupling
($g_{Z'}\sim 10^{-4}-10^{-3}$) which can accommodate the muon $(g-2)$ anomaly and is still allowed by other experimental
constraints, we show that we can get correct relic density of dark matter for wide range of dark matter mass 
($M_1 \sim 10-100$ GeV), although the gauge coupling constant $g_{Z'}$ is small.  
This is due to the fact that the annihilation cross section of dark matter pair is enhanced by
$M_1^4/m_{Z'}^4$ in the processes $N_1 N_1 \to Z' Z'$ or $N_1 N_1 \to Z' H_2$.  We also consider the constraints from
direct detection, collider searches.  
\end{abstract}
\maketitle
\newpage
\section{Introduction}
\label{sec:intro}
About 27\% of the universe is composed of dark matter, but we do not know its nature yet.
We may, however, find a clue for the dark matter in other sector of the standard model (SM), such as 
neutrino sector. One example is the models where the neutrino masses are generated radiatively
with dark matter as an essential component~\cite{radiative_nu}.

In Ref.~\cite{Baek:2015}, we extended Ma's scotogenic model~\cite{Ma_model} so that the model has gauged $L_\mu -L_\tau$ symmetry.
In fact, three symmetries $L_e-L_\mu$, $L_e-L_\tau$, and $L_\mu-L_\tau$, where $L_i$ is
the lepton number associated with the flavor $i$, can be gauged without
the extension of the SM particle content\footnote{We will denote $L_\mu-L_\tau$ as just $\mu-\tau$ for notational
simplicity.}. The gauge anomaly cancels between different generations.
In that paper we demonstrated that the neutrino mass matrix has two-zero texture due to the gauge symmetry, making
the theory very predictive. Especially we predicted the neutrino masses have inverted hierarchy and the Dirac CP phase is
close to maximal ($\sim 270^\circ$).

In this paper we consider the dark matter phenomenology of the model. Especially we will show that
we can get correct dark matter relic abundance and explain the muon $(g-2)$ ($(g-2)_\mu$) anomaly at the same time.
According to~\cite{trident}, almost all the region which can explain $(g-2)_\mu$ is excluded by
the neutrino trident production in $U(1)_{\mu-\tau}$ model. However, the region for $Z'$ mass, $m_{Z'} \lesssim 400$ MeV,
and for the extra $U(1)$ gauge coupling, $g_{Z'}\sim 3 \times 10^{-4} -10^{-3}$, is still allowed and can accommodate 
$(g-2)_\mu$ anomaly. In this paper we concentrate on this region, since the current experimental results still show
3-4$\si$ deviation from the SM predictions.

The analysis in this paper is applicable to more general dark matter models with light $Z'$ gauge boson coupled to right-handed neutrinos
where the lightest right-handed neutrino is the dark matter candidate. For example, the inert doublet scalar in the Ma model
is irrelevant for our discussion on dark matter and we would get similar results with this paper if only the right-handed neutrinos
have similar structure.

This paper is organized as follows.
In Section~\ref{sec:model}, we briefly review our model in the prospect of dark matter phenomenology.
In Section~\ref{sec:num}, we show numerical results.
In Section~\ref{sec:con}, we conclude.

\section{The model}
\label{sec:model}

\begin{center}
\begin{table}[t]
\begin{tabular}{||c||c|c|c|c|c|c|c|c|c||c|c|c||}
\hline\hline
          &$L_e$ & $L_\mu$ & $L_\tau$ & $e_R^c$ & $\mu_R^c$ & $\tau_R^c$ &$N_e^c$ &$N_\mu^c$ &$N_\tau^c$  &~$\Phi$~  & ~$\eta$& ~$S$~   \\\hline \hline
$SU(2)_L$ & \multicolumn{3}{c|}{$\bm{2}$} & \multicolumn{3}{c|}{$\bm{1}$}&\multicolumn{3}{c||}{ $\bm{1}$} & $\bm{2}$&  $\bm{2}$ & $\bm{1}$  \\\hline 
$U(1)_Y$  & \multicolumn{3}{c|}{$-1/2$}   & \multicolumn{3}{c|}{$1$}    & \multicolumn{3}{c||}{$0$}     & $+1/2$    & $+1/2$   & $0$  \\\hline
$U(1)_{L_\mu - L_\tau}$ & $0$ & $+1$ & $-1$ & $0$ & $-1$ & $+1$ & $0$ & $-1$ & $+1$ & $0$ & $0$ & $+1$ \\ \hline 
$Z_2$ & \multicolumn{3}{c|}{$+$} & \multicolumn{3}{c|}{$+$} & \multicolumn{3}{c||}{$-$} & $+$& $-$& $+$ \\ \hline\hline
\end{tabular}
\caption{The particle content and the charge assignment under $SU(2)_L \times U(1)_Y \times U(1)_{L_\mu-L_\tau} \times Z_2$.}
\label{tab:charge}
\end{table}
\end{center}

The original Ma model~\cite{Ma_model} introduces right-handed neutrinos $N_i^c$ ($i=e,\mu,\tau$), and $SU(2)_L$-doublet scalar $\eta$,
both of which are odd under discrete symmetry $Z_2$. As a consequence the lightest state of them do not decay into 
the standard model  (SM) particles and can be a dark matter candidate. 
The Yukawa interactions involving $L, N^c, \eta$ fields in the original Ma models are given by
\bea
{\cal L} &=& -\frac{1}{2} M_{ij} N_i^c N_j^c -y_{ij} \Phi^\dagger L_i e_j^c+ f_{ij}  \eta \cdot L_i N_j^c,
\label{eq:Ma}
\eea
where $\Phi$ is the SM Higgs doublet field and $\eta \cdot L_i \equiv \epsilon^{ab} \eta_a L_{ib}$ in $SU(2)_L$ space.
The neutrino mass terms come from one one-loop diagrams involving both $N_i^c$ and $\eta$~\cite{Ma_model}.

To extend the Ma model to local $U(1)_{\mu-\tau}$ symmetry, we just need to introduce one additional scalar
particle $S$ charged under $U(1)_{\mu-\tau}$  to break the abelian symmetry spontaneously.
The particle content and the charge assignment under $SU(2)_L \times U(1)_Y \times U(1)_{\mu-\tau} \times Z_2$
are shown in Table~\ref{tab:charge}.

The new gauge interactions are dictated by the gauge covariant derivative to give
\bea
\Delta  {\cal L} = \sum_{\psi=l_L^f, e_R^f, N_R^f} g_{Z'} Q'_\psi \, \ol{\psi}  \gamma^\mu  Z'_\mu\psi,
\eea
where $f=\mu,\tau$.

Due to $U(1)_{\mu-\tau}$ symmetry all the terms in (\ref{eq:Ma}) are not allowed. And the Yukawa interaction and right-handed neutrino
mass terms become more restricted to be
\bea
{\cal L} &=& -\frac{1}{2} M_{ee} N_e^c N_e^c -\frac{1}{2} M_{\mu\tau} (N_\mu^c N_\tau^c + N_\tau^c N_\mu^c)  \nl
           &-& h_{e\mu} (N_e^c N_\mu^c + N_\mu^c N_e^c) S - h_{e\tau} (N_e^c N_\tau^c + N_\tau^c N_e^c) S^* \nl
           &+& \eta \cdot (f_e L_e N_e^c +f_\mu L_\mu N_\mu^c +f_\tau L_\tau N_\tau^c) \nl
           &-& \Phi^\dagger (y_e L_e e_R^c +y_\mu L_\mu \mu_R^c +y_\tau L_\tau \tau_R^c) \nl
           &+& h.c,
\label{eq:Yuk}
\eea
where  all the fermions are Weyl spinors.
After $S$ gets vev $v_S$ ($ \langle S \rangle =v_S/\sqrt{2}$), we can see that the mass matrix of the right-handed neutrinos
can be written as
\bea
M_R =
\left(
\begin{array}{ccc}
M_{ee} & \frac{1}{2} h_{e\mu} v_S & \frac{1}{2} h_{e\tau} v_S \\
 \frac{1}{2} h_{e\mu} v_S  & 0  & M_{\mu\tau} e^{i \theta_R}  \\
 \frac{1}{2} h_{e\tau} v_S  &  M_{\mu\tau} e^{i \theta_R} & 0
\end{array}
\right).
\eea
By appropriate phase rotation, we can make all the parameters real except the one in $(2,3)$-component 
for which we allow CP violating phase $\theta_R$.
 The matrix $M_R$ is symmetric and can be diagonalized by a unitary matrix
\bea
 V^T M_R V = {\rm diag}(M_1, M_2, M_3).
\eea

The scalar potential of $\Phi$, $\eta$, and $S$ is given by
\bea
V &=& \mu_\Phi^2 |\Phi|^2 + \mu_\eta^2 |\eta|^2 + \mu_S^2 |S|^2 \nl
   &+& {1 \over 2} \la_1 |\Phi|^4  + {1 \over 2} \la_2 |\eta|^4 + \la_3 |\Phi|^2 |\eta|^2 + \la_4 |\Phi^\dagger \eta|^4
   +{1 \over 2} \la_5 \Big[ (\Phi^\dagger \eta)^2 + h.c. \Big] \nl
&+& {1 \over 2} \la_6 |S|^4 + \la_7 |\Phi|^2 |S|^2 + \la_8 |\eta|^2 |S|^2.
\eea
After $\Phi$ and $S$ get vev, $v$ and $v_S$, respectively, we can write 
\bea
\Phi = \left(\begin{array}{c}
0 \\
{1 \over \sqrt{2}} (v+h)
\end{array}
\right), \qquad
S = {1 \over \sqrt{2}} ( v_S + s),
\eea
in the unitary gauge. Then the two neutral states $h$ and $s$ can mix with each other with mixing angle $\alpha$, whose
mass eigenstates we denote as $H_1$ and $H_2$ with masses $m_{H_1}$ and $m_{H_2}$, respectively~\cite{Baek:2011aa}.
Here $H_1$ is the SM-like Higgs boson with $m_{H_1} \approx 125$ GeV.
In this paper we will assume this ``Higgs portal'' term, {\it i.e.} the $\la_7$, is small, because its mixing
angle is strongly suppressed by the study of Higgs signal strength~\cite{Baek:2011aa}.


\section{Muon $(g-2)$, relic density, direct detection of dark matter, and other tests of the model}
\label{sec:num}

In this section we concentrate on the dark matter phenomenology, especially the
relic density and the direct detection, of the model in
the region which can explain the muon $(g-2)$ anomaly.
Let us first consider the muon $(g-2)$ in our model.
The discrepancy between experimental measurement~\cite{Bennett:2004pv} and the SM prediction~\cite{Miller:2007kk}
\bea
 \De a_\mu \equiv a_\mu^{\rm exp} - a_\mu^{\rm SM} = (295 \pm 88) \times 10^{-11}, 
\eea
is about $3.4\si$ and can be explained by the $U(1)_{\mu-\tau}$ gauge boson contribution~\cite{Foot:1994vd,Baek:2001kca}.
Although the neutrino trident production process disfavors the $Z'$ explanation of muon $(g-2)$
for $m_{Z'} \gtrsim 0.4$ GeV~\cite{trident}, the light $Z'$ region is still consistent with $(g-2)_\mu$.

According to the Ref.~\cite{trident}, the allowed region for $(g-2)_\mu$ is characterized by
light $Z'$, $m_{Z'} \lesssim 0.4$ GeV and small $Z'$ gauge coupling constant, $10^{-4} \lesssim g_{Z'}  \lesssim 10^{-3}$.
For this small gauge coupling constant, it is naively expected the annihilation processes of the dark matter pair
at the electroweak scale dominated by~\cite{Baek:2008nz}
\bea
N_1 N_1 &\to& Z^{\prime *} \to l^+ l^-, \nu_l \ol{\nu}_l \quad (l =\mu, \tau), \nl
N_1 N_1 &\to& Z' Z',
\eea
would have very small cross sections. As a consequence, the dark matter relic density would overclose the universe.
It turns out that this is not the case.

\begin{figure}[tb]
\center
\includegraphics[width=0.8\textwidth]{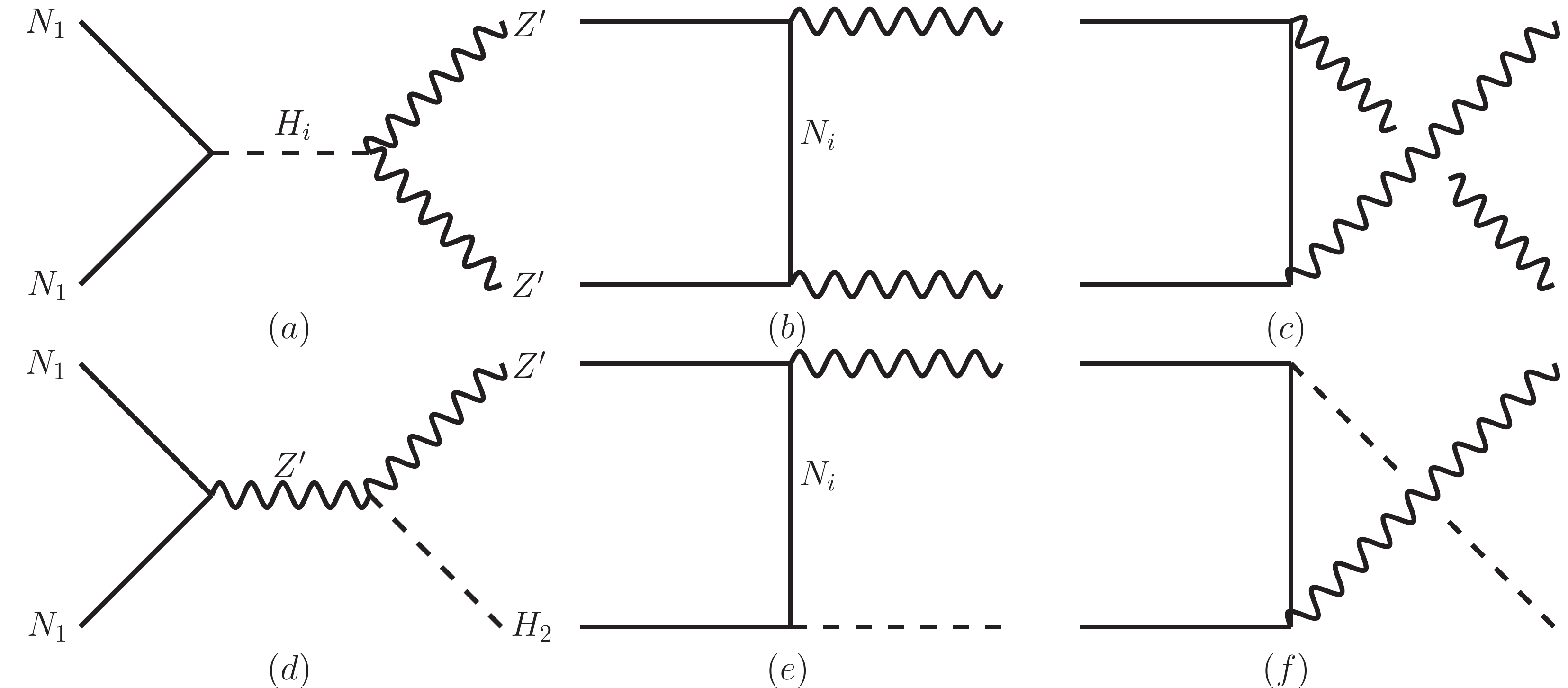}
\caption{Feynman diagrams for the processes, $N_1 N_1 \to Z' Z'$ and $N_1 N_1 \to Z' H_2$. Here
$H_i (i=1,2)$ are two scalar mass eigenstates and $N_i (i=1,2,3)$ are three right-handed neutrino mass
eigenstates.}
\label{fig:feynman}
\end{figure}

The dominant dark matter annihilation processes in our region of interest ({\it i.e.} light $Z'$ and small $g_{Z'}$) are
\bea
 N_1 N_1 \to Z' Z', \quad \text{and} \quad 
 N_1 N_1 \to Z' H_2, 
\eea
where $H_2$ is the lighter mass eigenstate between the SM Higgs and the $U(1)_{\mu-\tau}$  breaking scalar. 
For the second process to occur,  $H_2$ should also be light enough to be kinematically allowed.
The relevant diagrams are shown in Fig.~\ref{fig:feynman}.

We notice that the longitudinal $Z'$ polarization has enhancement factor, $\epsilon^{*\mu} (p) \sim p^\mu/m_{Z'}$, when its
energy is much larger than its mass. Since the total energy scale is almost fixed by the dark matter mass in dark matter
annihilation, there is an enhancement factor $M_1/m_{Z'}$ for each $Z'$ in the external or internal line in the annihilation
diagram. Consequently the diagrams with two $Z'$ gauge boson lines are most enhanced. And the enhancement factor in the
annihilation cross section is $M_1^4/m_{Z'}^4$. This large enhancement can compensate the suppression due to small gauge
coupling constant $g_{Z'}$ allowed by the $(g-2)_\mu$.  For example, explicit calculation shows the annihilation cross
section times relative velocity of the process, $N_1 N_1 \to Z' Z'$, in Fig.~\ref{fig:feynman} (a)-(c), is given by
\bea
\si v_{\rm rel} &\simeq& \frac{g_{Z'}^4 v_S^2 M_1^2 s}{4 \pi m_{Z'}^4} \Big[h_{e\mu} \Im(V_{11} V_{21}) +h_{e\tau} \Im(V_{11} V_{31})\Big]^2
\left(\frac{s_\al^2}{s-m_{H_1}^2}+ \frac{c_\al^2}{s-m_{H_2}^2}\right)^2\nl
&+& \frac{g_{Z'}^4 M_1^2 v_{\rm rel}^2}{12 \pi m_{Z'}^4}  \left(|V_{21}|^2-|V_{31}|^2\right)^4 \nl
&-& \frac{ \sqrt{2} g_{Z'}^4 v_S c_\al^2 M_1 s v_{\rm rel}^2}{24 \pi m_{Z'}^4 (s-m_{H_2}^2)} 
\left(|V_{21}|^2-|V_{31}|^2\right)^2 \Big[h_{e\mu} \Im(V_{11} V_{21}) +h_{e\tau} \Im(V_{11} V_{31})\Big] \nl
&+& \frac{ g_{Z'}^4 M_1^2 }{\pi m_{Z'}^4}  \sum_{j=2,3} \Bigg\{  \frac{2 M_1^2 M_j^2}{(M_1^2+M_j^2)^2}  \Big[
 \Im(V_{21}^* V_{2j} - V_{31}^* V_{3j})   \Re(V_{21}^* V_{2j} - V_{31}^* V_{3j}) \Big]^2 \nl
&+&  \frac{ c_\al^2 v_S  M_j s}{\sqrt{2}  (M_1^2+M_j^2) (s-m_{H_2}^2)} 
 \Im(V_{21}^* V_{2j} - V_{31}^* V_{3j})   \Re(V_{21}^* V_{2j} - V_{31}^* V_{3j}) \times \nl
&& \Big[h_{e\mu} \Im(V_{11} V_{21}) +h_{e\tau} \Im(V_{11} V_{31})\Big] \Bigg\},
\label{eq:sigv}
\eea
where $s=4 M_1^2/(1-v_{\rm rel}^2/4)$, $s_\al=\sin\al~(c_\al=\cos\al)$, and we show only the leading terms in $v_{\rm rel}$ and $M_1/m_{Z'}$.  
The vev of $S$ can be replaced by the $m_{Z'}$ using $v_S = m_{Z'}/g_{Z'}$.
Near the resonance region, {\it i.e.} $m_{H_i} \approx 2 M_1$,
the propagator, $1/(s-m_{H_i}^2)$, should be appropriately replaced by the Breit-Wigner form, $1/(s-m_{H_i}^2 + i
m_{H_i} \Ga_{H_i})$.
The 1st line results from Fig.~\ref{fig:feynman} (a), the 2nd line from $N_1$ contribution of
Fig.~\ref{fig:feynman} (b-c), and the 3rd line is the interference term between them.
The 4th line comes from $N_{2,3}$ contribution of Fig.~\ref{fig:feynman} (b-c), whose interference term
with Fig.~\ref{fig:feynman} (a) is the last term.
We assume the mixing angle $\al$ in the scalar sector is small, and we suppressed terms with $s_\al$ from the 2nd line
on. As can be seen clearly in (\ref{eq:sigv}), the $\si v_{\rm rel}$ has enhancement factor $M_1^4/m_{Z'}^4$ compared to
naive estimate which is given by $\sigma v_{\rm rel} \sim g_{Z'}^4/M_1^2$.
For the electroweak scale $N_1$ and $m_{Z'} \sim 100$ MeV, the enhancement factor can be of order $10^{12}$,
which can compensate the suppression due to $g_{Z'}^4 \sim 10^{-12}$, to give the correct relic density.

\begin{figure}[tb]
\center
\includegraphics[width=0.45\textwidth]{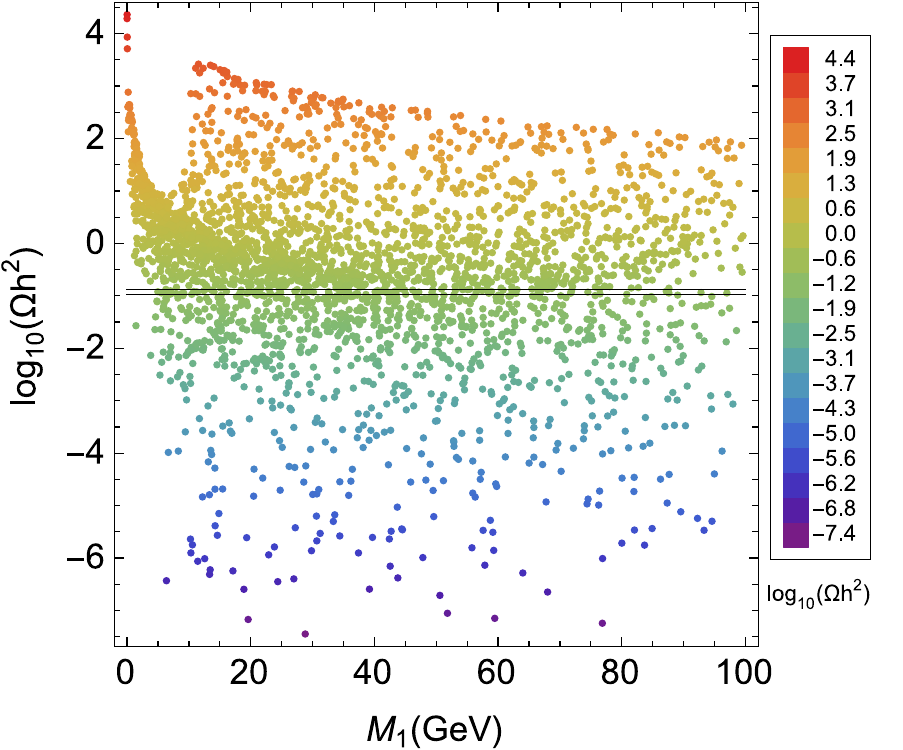}
\includegraphics[width=0.45\textwidth]{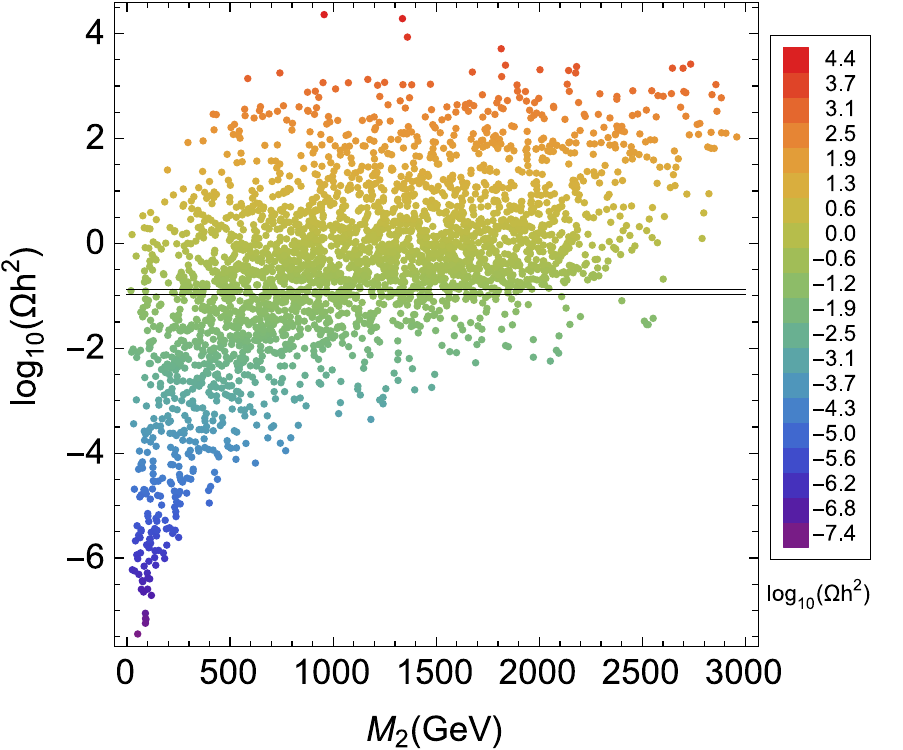}
\caption{The relic density versus $M_1$ (left panel) and $M_2$ (right panel). The horizontal lines represent $\pm 5 \si$
 values of Planck result, $\Om h^2 =  0.1199 \pm 0.0027$.}
\label{fig:Om}
\end{figure}

\begin{figure}[tb]
\center
\includegraphics[width=0.45\textwidth]{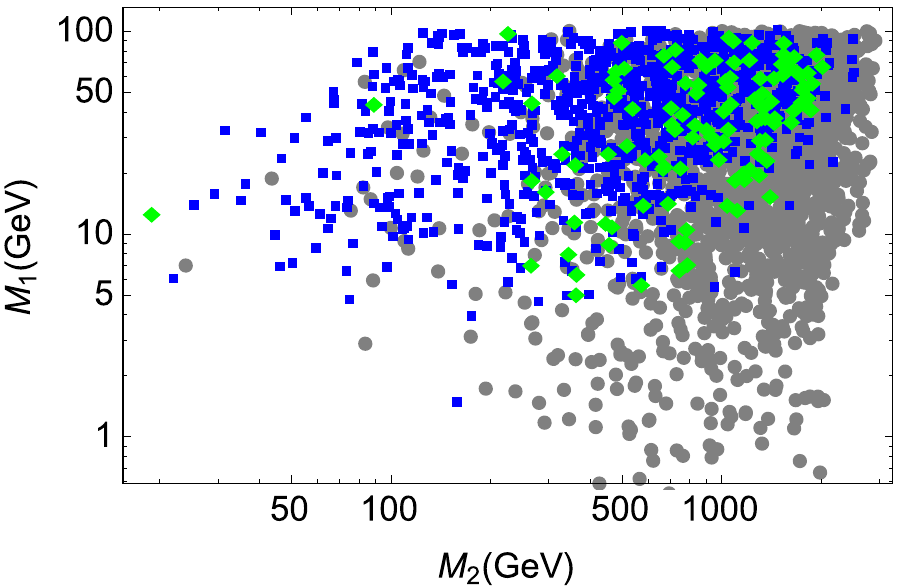}
\includegraphics[width=0.45\textwidth]{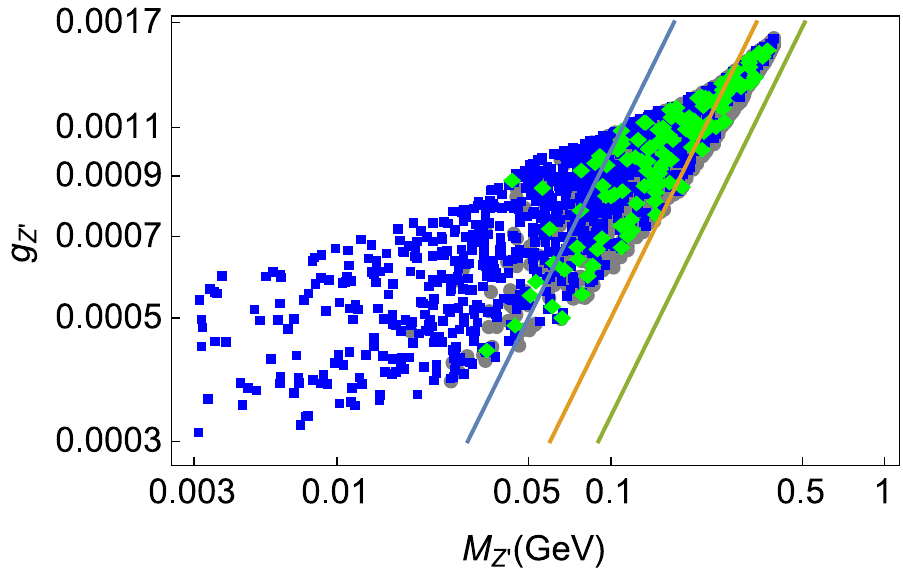}
\caption{Scatter plots in $(M_2,M_1)$ plane (left panel) and $(m_{Z'},g_{Z'})$ plane (right panel). All the points can
explain the $(g-2)_\mu$ at 2$\si$ level. The green points satisfy $0.1 < \Om h^2 <0.14$, the blue points $\Om h^2
<0.1$, and the gray points $\Om h^2 >0.14$. In the right panel the straight lines correspond to $m_{Z'}/g_{Z'}=100, 200,
300$ GeV from the left.}
\label{fig:scat}
\end{figure}

We scanned the region which can explain muon $(g-2)$ anomaly in $(m_{Z'},g_{Z'})$ plane~\cite{trident}, which can also
be seen in the right panel of Fig.~\ref{fig:scat}. For other parameters, we set 
\bea
 \al &=& 10^{-7}, \nl
m_{H_1} &=& 125~{\rm GeV}, \nl
\la_2 &=& \la_3 = \la_8 = 1, \nl
 m_{\eta^\pm} &=& m_{\eta_R} = m_{\eta_I} = 10~{\rm TeV},
\eea
where  $m_{\eta^\pm}$ and $m_{\eta_R(I)}$ are charged- and neutral-masses from inert scalar doublet $\eta$~\footnote{The
neutrino masses are sensitive to Yukawa couplings $f_i (i=e,\mu,\tau)$ in (\ref{eq:Yuk}) and are not strongly correlated
with the dark matter phenomenology}.
The change of the above parameters does not change our results much.
And  we scanned in the range
\bea
0 < &m_{H_2} &< \sqrt{4 \pi} m_{Z'}/g_{Z'} \nl
 10~{\rm GeV} < &M_{ee}, M_{\mu\tau}& < 100~{\rm GeV} \nl
 -4 \pi < &h_{e\mu}, h_{e\tau}& < 4 \pi \nl
 -\pi < &\theta_R& < \pi,
\eea
where used $m_{H_2} \approx \sqrt{\la_6} v_S$ to set the maximum value of $m_{H_2}$. 
With this scan, we get $M_1 \lesssim 100$ GeV and $M_2 \lesssim 3000$ GeV as we can see in Fig.~\ref{fig:Om}.

Fig.~\ref{fig:Om} shows the relic density versus $M_1$ (left panel) and $M_2$ (right panel). The horizontal lines
represent $\pm 5 \si$  values of Planck result, $\Om h^2 =  0.1199 \pm 0.0027$~\cite{Planck2013}.
We can see that the current relic density can be explained for wide range of dark matter mass, $M_1 \gtrsim 5$ GeV (See
also the left figure in Fig.~\ref{fig:scat}). We can also see that the $t-$channel $N_2$ contribution which is not
suppressed by $v_{\rm rel}^2$ can be important if it is not too heavy.

In Fig.~\ref{fig:scat}, we show scatter plots in $(M_2,M_1)$ plane (left panel) and $(m_{Z'},g_{Z'})$ plane (right
panel). All the points can explain the $(g-2)_\mu$ at 2$\si$ level. The green points satisfy $0.1 < \Om h^2 <0.14$, the
blue points $\Om h^2 <0.1$, and the Gray points $\Om h^2 >0.14$. In the right panel the straight lines correspond to
$M_{Z'}/g_{Z'}=100, 200, 300$ GeV from the left. We can see that the relic abundance of our universe can be explained if
$N_1$ is not too light ({\it i.e.} if $M_1 \gtrsim 5$ GeV) and $N_2$ has electroweak scale mass. The right panel shows
that the correct relic density can be obtained if $Z'$ is not too light. If $Z'$ is too light, {\it i.e.} $m_{Z'}
\lesssim 40$ MeV, the annihilation cross section becomes too large and the relic density becomes too small.

Since $Z'$ does not couple to quarks directly, our model does not have tree-level diagram for the direct detection of
dark matter off nucleons.  At one-loop level, $Z'$ can mix with photon via virtual $ \ell^+ \ell^- ~ (\ell=\mu,\tau)$
pair production and annihilation diagrams. Through this mixing the dark matter can scatter off nucleons.  To estimate
the elastic scattering cross section for direct detection it is convenient to introduce effective operator~\cite{Bell:2014tta}
\bea
{\cal L}_{\rm eff} = \frac{1}{\La^2} (\ol{N}_1 \ga^\mu \ga_5 N_1) (\ol{\ell} \ga_\mu \ell),
\eea
where $\ell =\mu,\tau$. The cut-off scale $\La$ is approximately given by $\La=m_{Z'}/g_{Z'}$.  As can be seen in the
right panel of Fig.~\ref{fig:scat}, the cut-off scale is in the electroweak scale.  Due to Majorana nature of $N_1$, the
vector current $\ol{N}_1 \ga^\mu N_1$ vanishes identically. The elastic scattering, however, is $p-$wave and the cross
section
is suppressed by $v_{\rm rel}^2 \approx 10^{-6}$~\cite{Bell:2014tta}. 

If we did not consider the muon $(g-2)$, the $U(1)_{\mu-\tau}$ gauge boson is also viable in the heavier $m_{Z'}$ or
larger $g_{Z'}$ parameter region.  In this case the $Z'$ can be searched for at colliders through 4$\mu$, 2$\mu$2$\tau$,
4$\tau$ production processes or missing $E_T$ signals in association with 2$\mu$ or 2$\tau$
events~\cite{Baek:2008nz}. The parameter region with $m_{Z'} \sim {\cal O}(10)$ GeV and $g_{Z'} \gtrsim 0.1$ is already
sensitive~\cite{Nojiri,trident} to LHC searches, $Z \to 4\mu$~\cite{CMS:2012bw,Aad:2014wra}. In the on-going LHC Run II
experiment wider region of parameter space will be covered~\cite{delAguila:2014soa}. The region of our interest, {\it
i.e.}, $g_{Z'} \sim {\cal O}(10^{-4})$ and $m_{Z'} \sim {\cal O}(100)$ MeV, may be searched for with dedicated study of
specific topology of events including the one such as lepton jet~\cite{trident}. This low $m_{Z'}$ would be tested better at
future high luminosity colliders such as FCC at CERN, Belle II, or planned neutrino facility LBNE.

The large $\nu_\mu$ flux from the dark matter annihilation at the galactic center can also be a signal of our
model~\cite{Baek:2008nz}. Those neutrinos can give additional contributions to the upward-going muon signals at the
Super-Kamiokande.  Although the photons emitted from the muons could contribute to the gamma rays from the galactic
center, the cross section turns out to be too small to explain the possible excess of gamma ray events from the
Fermi-LAT~\cite{Calore:2014xka}.

\section{Conclusions}
\label{sec:con}

In this paper we considered dark matter phenomenology of right-handed neutrino dark matter candidate in an extension of
Ma's scotogenic model with $U(1)_{\mu-\tau}$ gauge symmetry. We showed that we can explain the correct relic density of
dark matter and the anomaly of muon $(g-2)$ at the same time. We need light $Z'$ ($ m_{Z'} \lesssim 400$ MeV) and small
$U(1)_{\mu-\tau}$ gauge coupling ($3 \times 10^{-4} \lesssim g_{Z'} \lesssim 10^{-3}$). Although the gauge coupling
constant is small we showed that the longitudinal polarization of $Z'$ gauge boson in $N_1 N_1 \to Z' Z'$ annihilation
process can give large enhancement factor $M_1^4/m_{Z'}^4$ to get the correct relic abundance of dark matter.
Our model is not strongly constrained by the direct detection experiments of dark matter. 
However, the $Z'$ gauge boson can be searched for at the current LHC Run II and future high luminosity hadron or
neutrino collider experiments.

\begin{acknowledgments}
The author is grateful to Hiroshi Okada for useful discussion and carefully reading the manuscript. He also acknowledges
Korea Future Collider Study Group (KFCSG) for motivating him to proceed this this work. This work is supported in part
by National Research Foundation of Korea (NRF) Research Grant NRF-2015R1A2A1A05001869.
\end{acknowledgments}


\end{document}